\documentclass[conference]{IEEEtran}
\IEEEoverridecommandlockouts
\usepackage{cite}
\usepackage{amsmath,amssymb,amsfonts}
\usepackage{algorithmic}
\usepackage{graphicx}
\usepackage{bm}
\usepackage{textcomp}
\usepackage{xcolor}
\usepackage{caption}
\usepackage{subcaption}
\def\BibTeX{{\rm B\kern-.05em{\sc i\kern-.025em b}\kern-.08em
    T\kern-.1667em\lower.7ex\hbox{E}\kern-.125emX}}
\begin{document}

\title{Polarization Detection on Social Networks: dual contrastive objectives for Self-supervision}

\author{\IEEEauthorblockN{Hang Cui}
\IEEEauthorblockA{\textit{University of Illinois, Urbana Champaign} \\
hangcui2@illinois.edu}
\and
\IEEEauthorblockN{Tarek Abdelzaher}
\IEEEauthorblockA{\textit{University of Illinois, Urbana Champaign} \\
zaher@illinois.edu}
}

\maketitle

\begin{abstract}
Echo chambers and online discourses have become prevalent social phenomena where communities engage in dramatic intra-group confirmations and inter-group hostility. Polarization detection is a rising research topic for detecting and identifying such polarized groups. Previous works on polarization detection primarily focus on hand-crafted features derived from dataset-specific characteristics and prior knowledge, which fail to generalize to other datasets. This paper proposes a unified self-supervised polarization detection framework, which outperforms previous methods in both unsupervised and semi-supervised polarization detection tasks on various publicly available datasets. Our framework utilizes a dual contrastive objective (DocTra): (1). interaction-level: to contrast between node interactions to extract critical features on interaction patterns, and (2). feature-level: to contrast extracted polarized and invariant features to encourage feature decoupling. Our experiments extensively evaluate our methods again 7 baselines on 7 public datasets, demonstrating $5\%-10\%$ performance improvements.
\end{abstract}

\section{Introduction}
Polarization and echo chambers are common social phenomena where users tend to engage with online content that aligns with their preferred views. Social network platforms further diversify users' information exposure, which is often hyper-partisan and filled with polarizing biases. Polarization study is thus a new and promising research domain, usually considered self-supervised or unsupervised due to the sheer amount of online data. The problem has been studied qualitatively in areas such as social science and political silence~\cite{lai2020multilingual,garimella2017long}, and analyzed quantitatively in computer science literature~\cite{cinus2022effect,dash2022divided,ebeling2022analysis,sarmiento2022identifying,saveski2022perspective,ding2023same,efstratiou2023non,mok2023echo,cui2018recursive,cui2019semi,cui2021senselens,cui2024unsupervised,dou2023soft,dou2022empowering,peng2022self}. Examples include polarization detection, evolutions of polarization, and polarization reduction.

\begin{figure}
 \includegraphics[width=2\linewidth,trim = 2cm 7cm 10cm 3cm, clip]{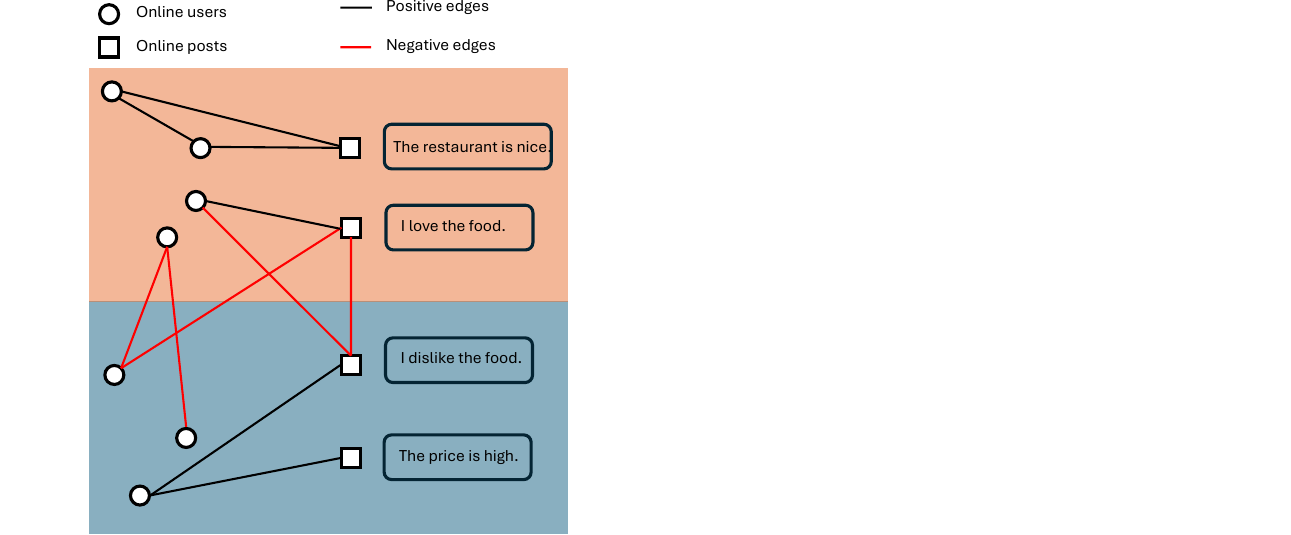}
 \caption{Toy example of a polarization detection task: The input consists of \textbf{up to} 3 types of edges (user-to-user, user-to-post, and post-to-post) of \textbf{up to} 2 types of signs (positive and negative).}
 \label{fig:toy}
\end{figure}

The polarization detection problem aims to identify and extract polarized groups from a given dataset. State-of-the-art solutions extract a set of features of high polarized characteristics~\cite{musco2018minimizing,yang2020hierarchical,darwish2020unsupervised,fang2022polarized,tu2022viral}, such as intra-group confirmation (also known as graph homophily or echo chamber), inter-group hostility, community wellness, and polarized frames (representative keywords and phrases) ~\cite{upadhyaya2023multi,ding2023same,lai2018stance}. Despite numerous attempts, previous methods either require sufficient labeled information or rely on handcrafted features derived from dataset characteristics. For example, \cite{efstratiou2023non,monti2023evidence} sololy extract hostile/toxic interactions across polarized groups. However, their studies also indicate that hostile interactions are not universal in all datasets.

A toy example of a polarization detection task is shown in fig.~\ref{fig:toy}. The input may consist of \textbf{up to} 3 types of edges (user-to-user, user-to-post, and post-to-post) of \textbf{up to} 2 types of signs (positive and negative). This paper proposes a unified self-supervision and fine-tuning objective working with various datasets of any combinations of input edge types and edge signs. 

\begin{figure*}
\centering
\begin{subfigure}{.5\textwidth}
  \centering
  \includegraphics[width=2\linewidth,trim = 1cm 13cm 10cm 2cm, clip]{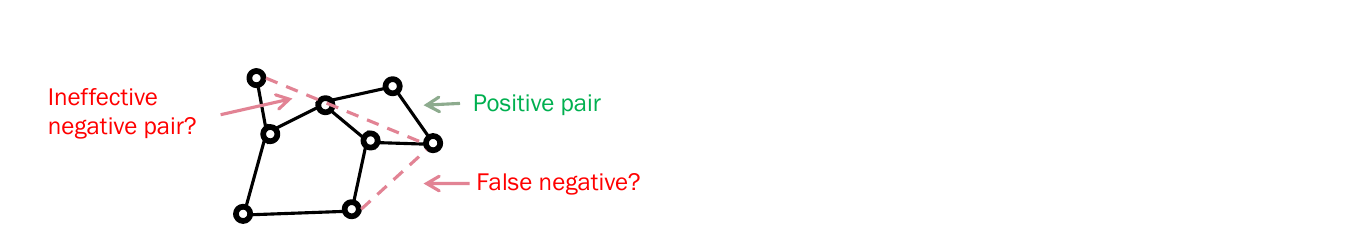}
  \caption{Interaction-level contrastive objective}
  \label{fig:obj1}
\end{subfigure}%
\begin{subfigure}{.43\textwidth}
  \centering
  \includegraphics[width=2\linewidth,trim = 3cm 13cm 10cm 2cm, clip]{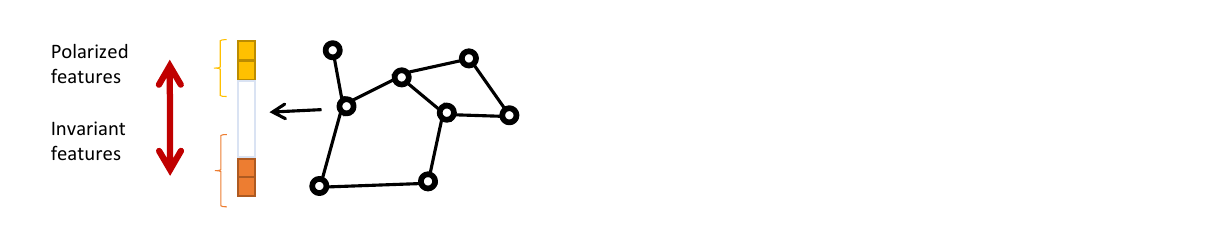}
  \caption{Feature-level contrastive objective}
  \label{fig:obj2}
\end{subfigure}
\caption{Dual contrastive objectives for self-supervised polarization detection: (a). contrast between positive interactions (what the user interacts with) and sampled `negative' interactions (what the user does not interact with). The red dashed lines present the possible sampled `negative' interactions. The key challenge is to eliminate false negatives and ineffective negative pairs. (b). contrast between polarized and invariant features.}
\label{fig:dual}
\end{figure*}

Our methods are based on two key observations of online discourses on social networks and polarization detection tasks. First, online discourses show a strong discrepancy in interaction patterns. For example, graph homophily methods maximize intra-group interactions, resembling the echo chamber phenomena~\cite{darwish2020unsupervised,fang2022polarized,tu2022viral}; other studies focus on maximizing inter-group hostility derived as the ratio of hostile interactions across and within polarized groups~\cite{efstratiou2023non,monti2023evidence}. Both examples can be understood as measuring the deviation of inter-group and intra-group interaction behaviors, which inspired us to the first objective, \textbf{interaction-level contrastive objective}, aiming to contrast between positive and negative examples of interactions. 

A naive approach is to sample supportive edges (such as likes and positive replies) as positive interactions and negative edges (such as hostile/toxic interactions) as negative interactions. However, supportive and hostile edges are not universally abundant in datasets. For example, political polarization on Reddit~\cite{efstratiou2023non} is shown as universally hostile, whereas tribalism (positive interactions within groups) is not universally observed. Whereas, political polarization on Twitter~\cite{dash2022divided} and many online discourses~\cite{ebeling2022analysis}, are opposite, with considerable intra-group confirmations but few inter-group hostilities. \textit{In short, both positive and negative interactions do not universally exist and are often imbalanced.} 

To counter the challenge, we propose a novel contrastive sampling framework that samples effective contrastive pairs requiring \textbf{only positive or negative} interactions. The key idea is to \textbf{contrast what a user supports/against, and what the user does not support/against}, as shown in fig~\ref{fig:obj1}. To rule out false negatives and ineffective negative pairs, we introduce a novel term, called \textit{polarization-induced silence}, which represents the lack of interaction due to polarization reasons (induced from \textit{polarized features}). \textit{Polarization-induced silence} are then contrasted with the observed positive/negative interactions to extract the high-quality decoupled features governing the interaction deviations. Another key benefit of the proposed framework is the \textbf{invariance} to edge types and signs: the sampled negative interactions are tailored to each observed positive edge and thus can be easily applied to any edge types and signs.

Second, extracted node features from online discourse demonstrate the decoupling of \textit{polarized features} and \textit{invariant features}: online interactions (often known as engagements) are determined by both polarization-related features and polarization-invariant features. For example, an online user tends to engage with local topics, although the locality feature is not polarized. In addition, various topics possess different levels of background engagement. For example, political communities interact significantly more (both positively and negatively) than in tourism/gaming communities. We show that both \textit{invariant features} and \textit{invariant features} are essential in extracting fine-grained features describing the polarization phenomena. Therefore, the second objective, \textbf{feature-level contrastive objective}, is designed to encourage decoupling polarized and invariant features.

In addition, we propose a \textit{unified polarization index} to measure the polarization level given a raw dataset. Our method is functionally unsupervised but is robust to various supervised signals and datasets. Our contribution includes:
\begin{enumerate}
    \item A novel dual contrastive objective (DocTra) for polarization detection and clustering/classification. Our method requires no prior knowledge or hand-crafted methods, is flexible to supervised signals, and robust to various noises.
    \item A novel unified polarization index able to well distinguish polarized graphs and unpolarized graphs.
    \item Extensive experiment demonstrates the effectiveness of our method.
\end{enumerate}

\section{Related Works}
\subsection{Polarization}
Online users tend to consume content that aligns with their personal beliefs, resulting in the polarization phenomenon. Polarization is further intensified by filter bubbles~\cite{chitra2020analyzing} (such as recommender algorithms) and online discourses~\cite{mok2023echo}. Recently, polarization has been extensively studied in the research literature, including political science\cite{barber2015causes}, social science~\cite{levin2021dynamics}, and computer science\cite{zhou2024modeling,musco2018minimizing,racz2023towards}.

Polarization detection is a fundamental problem that aims to detect and classify (cluster) related polarized nodes within an input graph. Previous attempts mostly focus on identifying polarization-related characteristics within the input dataset via handcrafted models and graph self-supervised learning.

Most previous methods utilize the graph structure to extract polarized features. Early models are based on the famous \textit{Friedkin–Johnsen opinion formation model}\cite{zhou2024modeling,musco2018minimizing,racz2023towards}, which is essentially a non-learnable message passing model. The latter methods utilize random walks\cite{adriaens2023minimizing}, variational graph encoder\cite{li2021unsupervised}, and polarized graph neural networks\cite{fang2022polarized} to generate polarized embeddings. Other works explore dataset-specific characteristics. For example, \cite{efstratiou2023non} extracts hostile/toxic interactions across polarized groups. \cite{sarmiento2022identifying} proposes several key network characteristics, including the number of unique tweets, retweet relations, and user similarities.

Other methods exploit text features using fine-tuned large language models, including BERT\cite{ding2023same}, emotional stance\cite{ding2023same}, sentence transformer\cite{chaturvedi2024bridging}, topic modeling\cite{chaturvedi2024bridging}, and universal sentence encoder\cite{dash2022divided}. However, linguistic-based methods require \textbf{substantial} prior knowledge to fine-tune the pre-trained language models.

Our method follows the \textbf{structure-based} method, supplemented with \textbf{linguistic-based} methods as \textit{optional} supervised signals, where some nodes can be evaluated using linguistic encoders into labels. The benefit of such assumptions is two-folded: (1). The structure-based approaches can be widely deployed to real-world datasets without prior knowledge and supervision. (2). Linguistic-based methods often provide valuable labeled data facilitating initial classification/clustering.

\subsection{Graph Contrastive Learning}
Graph contrastive learning~\cite{hou2023graphmae2,Zhu:2020vf,zhang2021canonical} is a popular pre-training objective in graph self-supervised learning, where the graph/node representations are pre-trained unsupervised on the contrastive objective prior to the downstream tasks. The key principle is to preserve the pre-trained representation against the augmented views of the original input. Most previous works use graph corruption as the augmented view: the original graph is corrupted via edge dropping, feature masking, and node removal. The corrupted graph is then encoded and contrasted with the original graph on node-level and graph-level objectives. The optimal choice of augmentation methods often depends on the downstream tasks, where the augmentation methods can decouple spurious features while keeping the task-dependent features intact~\cite{wen2021toward,xu2021infogcl}.

To the best of our knowledge, both interaction-level contrastive objective and feature-level contrastive objective are novel in graph contrastive learning. Both objectives are tailored for polarization detection tasks and are flexible on various types of inputs and supervisions.

\section{The Polarization Detection Problem}

Given an attributed graph $G(V,E,X)$, where $V,E,X$ are node sets, edge sets, and input features; the objective is to detect polarized groups (classes) $C$ and classify/cluster the related nodes into the groups. Following previous literature, we consider \textbf{binary} polarization detection task, such that $|C|=2$, because (1). most of the public datasets and real-world controversies are binary, such as political parties (Republican vs Democrat), support/against stances on a controversial topic (COVID vaccination stance); (2). multi-party polarization detection tasks can be reduced to multiple binary polarization detection tasks.

The nodes $V$ can be online users or online posts (denoted as items). The input feature matrix $\bm{X}$ is usually pre-obtained from encoding the users and items via a linguistic encoder. For example, in Reddit datasets, items are the threads under which users post and reply. In Twitter and Facebook datasets, we follow the previous practice of clustering highly similar posts into items to reduce sparsity~\cite{dash2022divided}. Since there can be two types of nodes (users and items), the input graphs are either homogeneous (one type of nodes) or heterogeneous (of two types of nodes). 

Since most datasets do not provide edge signs, the edge set is unsigned by default (where only positive or negative edges are available) for generalization purposes. However, our method can be easily extended to signed graphs. Without loss of generality, the following sections consider edges as bipartite interactions between users and items (for example, a user reposts, likes, or replies to an online item) for discussion purposes since it is the most common interaction on social networks. Note that, our method can be equally applied to unipartite interactions: user-to-user interaction and item-to-item interaction.

The polarized classes $C$ are assumed unknown. This paper uses soft group(class) assignment of assignment matrix $\bm{R}$, such that $R_{:1}+R_{:2}=1$. In addition, we denote the embedding matrix as $\bm{H}$, polarized related terms using superscript $po$, and invariant terms using superscript $in$. For example, \textit{polarized features} are characterized as embedding matrix $\bm{H}^{po}$ whereas \textit{invariant features} as embedding matrix $\bm{H}^{in}$, with $\bm{H} = \bm{H}^{po}\mathbin\Vert \bm{H}^{in}$, $\mathbin\Vert$ denotes concatenation.

\subsection{Key Discrepancy to General Node Classification Problems}
The above problem formulation is similar to the general-purpose node classification problem. We emphasize two key differences:
\begin{itemize}
    \item The polarization detection problem is often unsupervised or extremely few-shot. Therefore, the proposed methods must effectively utilize the key characteristics of social discourse and polarization.
    \item The polarization datasets consist of input graphs with various characteristics and noises: (1). various network structures: polarization datasets consist of different edge densities (sparse to dense graphs) and edge types (positive and/or negative edges, bipartite edges and/or unipartite edges). (2). neutral nodes (nodes that do not belong to any classes) and irrelevant nodes (outlier nodes that are not relevant to the topic of interest).
\end{itemize}

Our proposed method is flexible in various input graphs in a unified framework without any pre-assumed labels, and also effectively integrates (\textbf{optional}) labeled information. This paper demonstrates two types of supervision: (1). Node labels: a subset of node $V_l$ can be pre-labeled of their polarized stance via a domain expert. (2). Class initialization: the unknown polarized class can often be initialized via topics obtained from topic models or online communities (such as Reddit (sub)-communities). 

\section{Motivation}
Previous works in polarization detection tasks suffer from two major weaknesses: (1). reliance on prior knowledge and hand-crafted features in both model design and dataset collection. (2). low robustness to various input characteristics and noise. This paper proposes (1). a unified self-training, fine-tuning framework tailored for polarization detection tasks with minimal or no pre-assumptions and handcrafted methods, (2). a polarization metric, measuring the polarization level of the input datasets, aiming to effectively distinguish between polarized and unpolarized datasets. 

Our method integrates two self-supervised objectives: 
\begin{itemize}
\item Interaction-level contrastive objective: Contrast positive and negative examples of interactions inspired by the deviation of interaction behaviors in online discourses, such as inter-class echo chambers and intra-class hostility.
\item Feature-level contrastive objective: Contrast \textit{polarization-specific characteristics}, namely \textit{polarized features}, and \textit{cross-class invariant features}, namely \textit{invariant features}, aiming to extract finer-grained features governing both polarized and unpolarized phenomenon.
\end{itemize}

We show that the above two objectives can be trained jointly in contrastive self-supervised learning, as shown in fig.~\ref{fig:framework}.

\subsection{Interaction-level contrastive objective}
Inspired by previous attempts at analyzing inter-group hostility and intra-group confirmations, we propose to train a \textbf{contrastive objective between positive and negative examples of interactions}. There are two major advantages of interaction-level contrastive objective:
\begin{itemize}
\item enables easy adaptation to various edge densities and edge types in polarization datasets.
\item reflects the interaction deviations between classes in online discourses.
\end{itemize}

A naive approach is to sample the positive/negative examples directly from the hostile/supportive interactions. However, the co-existence of both positive and negative interactions is not universally abundant across datasets. For example, political polarizations on Reddit~\cite{efstratiou2023non} are shown almost universally hostile, whereas political polarizations on Twitter~\cite{dash2022divided,ebeling2022analysis} are directly opposite, with considerably more intra-group positivity than inter-group hostilities. This imbalance of positive/negative interactions hinders the sampling of high-quality contrastive pairs.

To solve the above challenge, we propose a novel contrastive sampling method that \textit{only requires positive or negative interactions}. The key idea is to \textbf{contrast what a user supports/against, and what the user does not support/against}, which is often known as \textbf{silence} behavior in online interaction: why no edges(interactions) between node pairs). However, interpreting \textit{silence} is considerably more challenging than interpreting observed interactions, due to the unavailability of associated contents and various underlying reasons. For example, in the social network settings, no edges may arise from various potential reasons: the user might not observe the topic on social media; the user might abstain from interacting with it due to lack of engagement, or the user might disagree with the content due to polarized opinions; and so on.

Therefore, we focus on extracting \textbf{polarization-induced silence}, where the user \textit{silences} due to polarization-related features. We define \textit{polarization-induced silence} as the item that a user does not interact with but would otherwise likely interact with it without polarized stances. \textit{Polarization-induced silences} can be understood as the set of \textit{most similar silences}, aligned with the \textit{most effective} contrastive sampling strategy proposed in previous contrastive learning literature~\cite{wen2021toward,xu2021infogcl}. \textit{Polarization-induced silences} are then paired with the corresponding positive/negative interactions in the contrastive framework. 

\begin{figure*}
 \includegraphics[width=\linewidth,trim = 2cm 7cm 1cm 1cm, clip]{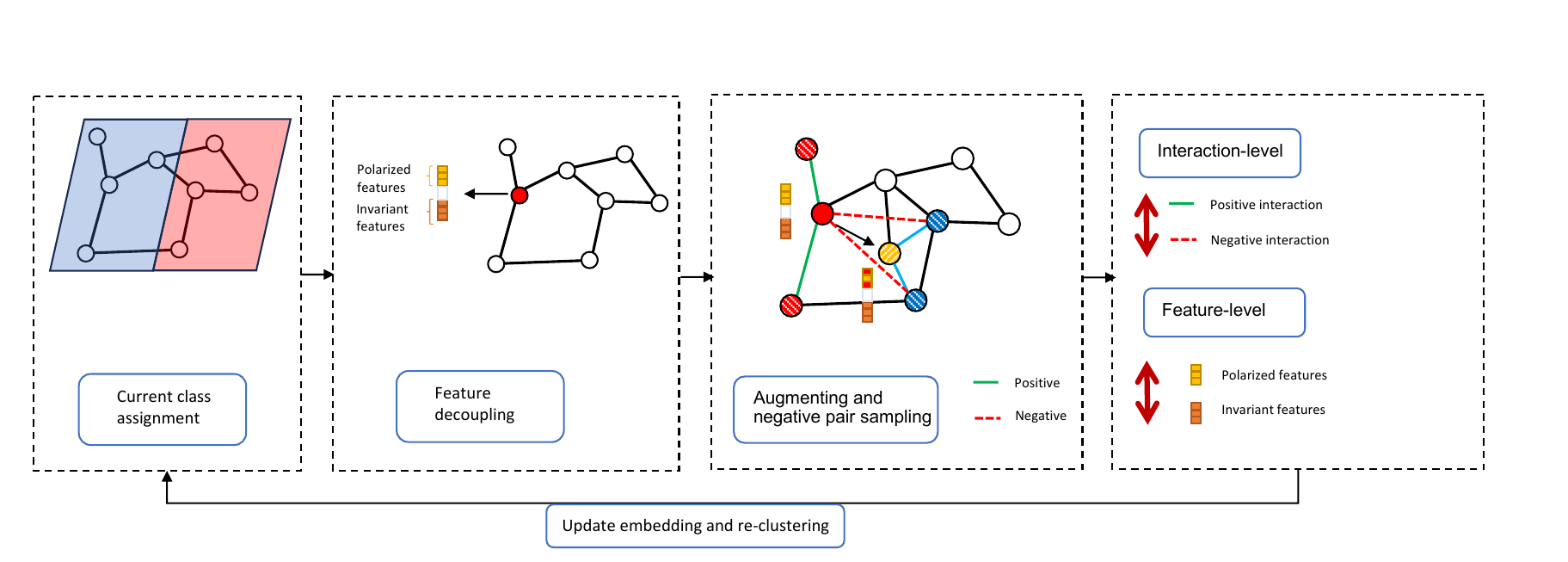}
 \caption{Iterative framework of DocTra: (a). obtain class assignments from current embeddings; (b). obtain decoupled features; (c). given the anchor node (red), sample positive interactions (green line) and negative interactions (red dash line) by augmenting the anchor node and solving eq.(1); (d). performing contrastive learning on both objectives to update the embeddings.}
 \label{fig:framework}
\end{figure*}

Formally, the polarized stance of node $i$ is characterized via extracted \textit{polarized features} $\bm{H}_i^{po}$. We then apply an learnable augmentation function $f()$, (by default feature perturbation) on $\bm{H}_i^{po}$, such that

\noindent\begin{align}
    \nonumber V_i^-=&\{j| Connect(\bm{H}_i^{po}||\bm{H}_i^{in},\bm{H}_j^{po}||\bm{H}_j^{in})<\sigma_1,\\ &Connect(f(\bm{H}_i^{po})||\bm{H}_i^{in},\bm{H}_j^{po}||\bm{H}_j^{in})>\sigma_2\}\\
    V_i^+=&\{j|j\in N_i\}
\end{align}
where $Connect(,)$ is a pre-trained (such as MLP) or pre-defined (such as inner product) link prediction model; $f()$ is an augmentation function; $\sigma_1, \sigma_2$ are hyperparameters for lower and upper link prediction scores; $N_i$ is the set of neighboring nodes of $i$. In simple words, the above formulation outputs node set $V_i^- = \{j\}$ of low($<\sigma_1$) connectivity to $i$ but high($>\sigma_2$) connectivity after augmenting \textit{polarized features}. The exact derivation of \textit{polarization-induced silence} will be introduced in the later section. For example in fig.~\ref{fig:framework}c, the anchor node (red) is augmented into the yellow node by augmenting polarized features with a learnable augmentation function. The two blue-shaded nodes are the polarization-induced silence nodes: the anchor node (red) does not interact with, but the augmented nodes would interact. Therefore, the two red-dashed interactions between the anchor node and the polarization-induced silence nodes are the sampled negative interactions for effective contrastive learning.

Given the positive and negative sets, we can then formulate the \textit{interaction-level contrastive objective}:
\begin{align}
    \mathcal{L}_i = \sum_{-\sim V_i^-,+\sim V_i^+}\frac{d_i(H^{po}_i,H^{po}_+)}{d_i(H^{po}_i,H^{po}_+)+d_i(H^{po}_i,H^{po}_-)}
\end{align}
where $d_i(,)$ is a distance metric measuring the node discrepancy. $\mathcal{L}_i$ contrasts the node discrepancy on \textit{polarized features} between positive and negative interaction samples.

\subsection{Feature-level contrastive objective}

Previous works usually extract \textit{polarized features} and \textit{invariant features} independently. We argue that both features are heavily intertwined in real-world interaction patterns. For example, an online user likely engages more local content, but the locality features might not be relevant to polarity. In addition, the underlying topics usually possess different background engagement levels. For example, online users in political communities interact significantly more (both positively and negatively) than in tourism/gaming communities. Such background engagement levels should be incorporated into polarization measurement. Thanks to the success of GNN-based methods, \textit{invariant features} can easily be extracted alongside the \textit{polarized features}. Formally, we employ parallel pairs of encoders: polarized encoder $enc^{po}$ and invariant encoder $enc^{in}$ to extract \textit{polarized features} $H^{po}$ and \textit{invariant features} $H^{in}$ respectively:
\begin{align}
    &H^{po} = enc^{po}(G,X)\\
    &H^{in} = enc^{in}(G,X)\\
    &\mathcal{L}_f = \sum_{i\neq j\in V}\frac{d_f(H_i^{po},H_j^{po})}{d_f(H_i^{in},H_j^{in})}
\end{align}
where $d_f(,)$ is a distance metric measuring the discrepancy of two feature vectors. $\mathcal{L}_f$ is the feature-level contrastive objective encouraging the decoupling of the two feature spaces.

Another benefit of decoupling \textit{polarized features} and \textit{invariant features} is to generate `hard' contrastive pairs for \textit{interaction-level contrastive objective}. `Hard' implies challenging contrastive pairs that are non-trivial to the current classifier/clustering, as suggested in studies of efficient contrastive learning. The exact formulation is shown in next section.

\section{Doctra}
The previous section introduced the dual contrastive objectives of our framework:

\begin{align}
    H^{po} =& enc^{po}(G,X)\\
    H^{in} =& enc^{in}(G,X)\\
    \label{V-}
    V_i^-=&\{j| Connect(\bm{H}_i^{po}||\bm{H}_i^{in},\bm{H}_j^{po}||\bm{H}_j^{in})<\sigma_1,\\ &Connect(f(\bm{H}_i^{po})||\bm{H}_i^{in},\bm{H}_j^{po}||\bm{H}_j^{in})>\sigma_2\}\\
    V_i^+=&\{j|j\in N_i\}\\
    \mathcal{L}_i =& \sum_{-\sim V_i^-,+\sim V_i^+}\frac{d_i(H^{po}_i,H^{po}_+)}{d_i(H^{po}_i,H^{po}_+)+d_i(H^{po}_i,H^{po}_-)}\\
    \mathcal{L}_f =& \sum_{i\neq j\in V}\frac{d_f(H_i^{po},H_j^{po})}{d_f(H_i^{in},H_j^{in})}\\
    max&\mathcal{L} = \mathcal{L}_i + \mathcal{L}_f
\end{align}

This section presents (1) an efficient solver for the dual objectives, (2) how to incorporate supervised signals, and (3) finally, a unified polarization index. 

\subsection{Efficient Solver for the Dual Contrastive Objective}
$enc^{po}$ and $enc^{in}$ are the graph encoders, common choices are GCN and GAT. $V_i^+$ is a straightforward sampling of neighboring nodes of $i$. Therefore, the challenging parts are (1). $V_i^-$ and (2). the joint training of $\mathcal{L}_i$ and $\mathcal{L}_f$.

$\bm{V_i^-}$. $V_i^-$ is the node set $\{j\}$ of low($<\sigma_1$) connectivity to $i$ but high($>\sigma_2$) connectivity after augmenting \textit{polarized features} $\bm{H}_j^{po}$ via an augmentation function $f$. The most popular feature-based augmentation functions are:
\begin{itemize}
\item perturbation: $f(h) = h + \mu, |\mu|<B$
\item interpolation: $f(h,h') = hx+bh', a+b=1$
\end{itemize}

With both $f()$, eq.(\ref{V-}) can be solved via gradient descent. However, this brute-force method is expensive as the gradient descent is applied to a parameterized link prediction model $Connect()$ on every node pair $i,j$. 
Inspired by the previous works on complexity reduction of neural networks\cite{mahabadi2022perfect}, $Connect(H_i,H_j)$ can be approximated via $M(H_i)\cdot M(H_j)$, where $M()$ is often called adaptors, which only takes a singular input. The key benefit of using the adaptors is that $M(H_i)$ is fixed for node $i$, and thus the gradient descent is only applied to $M(H_j)$. Although this formulation is cheaper than $Connect(H_i,H_j)$, this still requires $O(V^2)$ gradient descends.

To further simply the computation, we make the following relaxation:
\begin{align}
    M(H) = M(H^{po}||H^{in}) \sim M^{po}(H^{po})||M^{in}(H^{in}) 
\end{align}
such that the adaptors are applied to \textit{polarized features} and \textit{invariant features} independently. This is a reasonable relaxation as those two features are extracted separately using two graph encoders. The relaxation results in:
\begin{align}
    \nonumber M(H_i)\cdot M(H_j) &\sim [M^{po}(H_i^{po})\cdot M^{po}(H_j^{po})]\\
    &+[M^{in}(H_i^{in})\cdot M^{in}(H_j^{in})]
\end{align}
$M^{in}(H_i^{in})\cdot M^{in}(H_j^{in})$ is fixed throughout the epoch, and $M^{po}(H_i^{po})\cdot M^{po}(H_j^{po})$ is likely small since $i$ and $j$ are not connected. Therefore, we thresholds $\mathcal{M} = M^{in}(H_i^{in})\cdot M^{in}(H_j^{in})$, such that $\mathcal{M}_{ij}>\sigma_3$ to obtain the set $V_i^-$.\\

\noindent\textbf{Joint training $\bm{\mathcal{L}}$.} With $V_i^-$ and $V_i^+$, $\mathcal{L}$ can be trained iteratively on $H^{po}$ and $H^{in}$ by fixing the other. When trained unsupervised (self-supervised), the model must be carefully initialized. We utilize $\mathcal{L}_f$ along to initialize the embeddings, encouraging decoupled initializations of polarized and invariant features.\\

\begin{figure}
 \includegraphics[width=1.8\linewidth,trim = 4cm 11cm 10cm 3cm, clip]{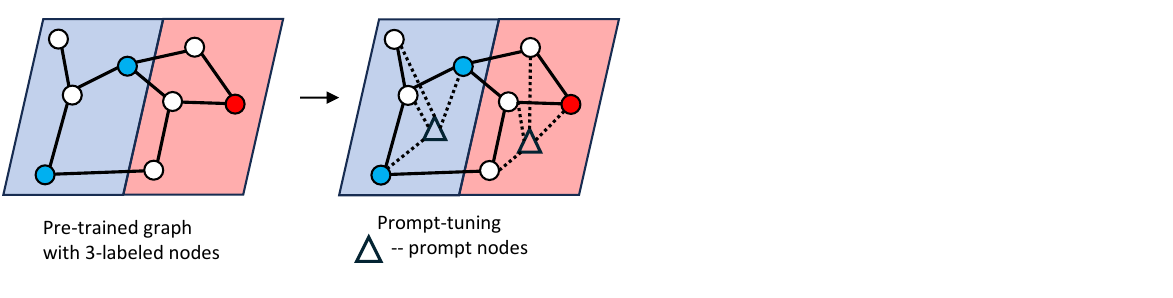}
 \caption{Prompt-tuning framework: the triangle nodes are the learnable prompt nodes added to the input graph. The dashed lines are the induced edges derived from Connect(,).}
 \label{fig:prompt}
\end{figure}

\noindent\textbf{Clustering}. After self-supervised learning, unsupervised clusters can be obtained from \textit{polarized} and \textit{invariant} features. The general idea is to apply soft clustering algorithms on \textit{polarized features} to obtain cluster centers and using \textit{invariant features} to filter out irrelevant nodes. This paper uses the standard soft k-means assignment~\cite{wilder2019end} on \textit{polarized features}:
\begin{align*}
    &r_{ik} = \frac{exp(-\beta\lVert H^{po}_i-\mu_k\rVert)}{\sum_l exp(-\beta\lVert H^{po}_i-\mu_l\rVert)}\\
    &\mu_k = \frac{\sum_ir_{ik}H^{po}_i}{\sum_ir_{ik}}
\end{align*}

\noindent\textbf{Irrelevant and neutral nodes}. Real-world datasets may consist of substantial irrelevant or neutral nodes that must be well-distinguished from the clustered polarized classes. Thanks to the decoupled features, we can identify both types of nodes via outlier detection methods:
\begin{itemize}
    \item irrelevant nodes denote the nodes out of the scope of interest to the topic. We propose to apply outlier detection on \textit{invariant features} (features shared across polarized classes). This paper uses the standard deviation (by default 2 standard deviations) of \textit{invariant features} to threshold the irrelevant nodes. 
    \item Neutral nodes denote the nodes that are indifferent to both polarized classes. We use the soft assignment to threshold the neutral nodes (by default $max_k r_{ik}<0.7$). 
\end{itemize}

\subsection{Incorporating Supervision via Semi-supervision}
Supervised signals are commonly available in real-world applications. Adaptation to supervision is, therefore, an important factor for graph learning methods. This paper considers two (\textbf{optional}) types of supervision: (1). Node labels: a subset of node $V_l$ is accurately pre-labeled of their polarized stance. (2). Class initialization: the polarized classes (groups) can often be (roughly) initialized by topic models or online communities (such as Reddit (sub)-communities).

\noindent\textbf{Node labels} are integrated in two ways:
\begin{itemize}
\item If the labels are abundant ($>5\%$), we can follow the previous graph self-supervised learning by freezing the node embeddings ($H$) and then training a logistic classifier to replace clustering.
\item If the labels are not abundant, we instead add a semi-supervised objective: $min \mathcal{L}_n = \sum_{l\in V_l} ||H_l^{po}-\mu_k||$, where $k$ is the labeled class of $l$.
\end{itemize}

\noindent\textbf{Class initialization} assumes an initial assignment matrix $R=\{r_{ik}\}$. To obtain the initial embedding, we employ an initialization objective (discarded after the first few epochs) encouraging the alignment of \textit{polarized features} towards the class center:
\begin{align}
    min\mathcal{L}_c = \sum_{i\in V}||H_i^{po}-\mu_k||
\end{align}
where $k$ is the initial class of $i$.

\subsection{Incorporating Supervision via Prompt-tuning}
Prompt-tuning is a well-applied method in natural language processing and computer vision tasks and has recently been adapted to graph tasks~\cite{sun2023graph}. The core idea is to freeze the pre-trained models and then add a set of learnable prompt parameters, which are updated during prompt-tuning.

The detailed model is shown in fig.~\ref{fig:prompt}. Thanks to our interaction-level contrastive objective, the prompt nodes can be effortlessly added to the input graphs.

\subsection{Unified Polarization Metric}

The most popular polarization metric on graph $G$ is the \textit{polarization-disagreement index} $I()$, which is the summation of \textit{polarization index} $P()$ and \textit{disagreement index} $D()$~\cite{zhou2024modeling}:
\begin{align}
&P(H) = Var(H)\\
&D(H) = \sum_{(i,j)\in E} w_{ij}d(H_i,H_j)\\
&I(H) = P(H) + D(H)
\end{align}
where $P(G)$ measures the variance of the feature matrix and $D(G)$ measures the sum of discrepancy along edges. $w_{ij}$ is an optional weight matrix.

The above index has two key weaknesses: (1) It does not consider the datasets' background engagement levels; (2) It does not consider the effect of outliers. We propose a simple modification to overcome the above two weaknesses. Our formulation is as follows:
\begin{align}
&P(H) = \frac{Var(H^{po})}{Var(H^{in})}\\
&D(H) = \sum_{(i,j)\in E} w_{ij}\frac{d(H^{po}_i,H^{po}_j)}{d(H^{in}_i,H^{in}_j)}\\
&I(H) = P(H) + D(H)
\end{align}

Our unified index (1). scales down the background engagement level via \textit{invariant features}. (2). reduces the effect of outliers since their $Var(H^{in})$ are large.
 
\section{Experiments}
The experiment session studies 3 research questions:
\begin{enumerate}
    \item Can our proposed Doctra method outperform baselines on polarization clustering?
    \item Can Doctra incorporate labeled information better than baselines?
    \item Can our unified polarization metric distinguish polarized graphs from unpolarized graphs?
\end{enumerate}

\subsection{Main Experiment}

\noindent\textbf{Datasets}. We include a variety of publicly available datasets used in previous polarization-related papers: Twitter datasets on political discourse~\cite{panda2023}, Chilean unrest~\cite{sarmiento2022identifying}, and COVID vaccine stance~\cite{nimmi2022pre}; Reddit dataset of r/news~\cite{baumgartner2020pushshift}; Wikipedia datasets on editor communication and election~\cite{rossi2015network}; other local social networks~\cite{rossi2015network}. The dataset statistics are shown in table.~\ref{tab:stats}

\noindent\textbf{Baselines}. We compare our method \textit{Doctra} with state-of-the-art self-supervised method: GraphMAE2~\cite{hou2023graphmae2}, Grace~\cite{Zhu:2020vf}, and CCA-SSG~\cite{zhang2021canonical}; general polarization-detection method: polarized graph neural networks~\cite{fang2022polarized}, variational graph encoder~\cite{li2021unsupervised}, and FJ model~\cite{matakos2017measuring}; characteristic-specific method on hostile interactions~\cite{efstratiou2023non} and on (re)tweet patterns~\cite{sarmiento2022identifying}.

\begin{table}[]
\caption {\% Dataset statistics}
\centering
\label{tab:stats}
\begin{tabular}{l|l|l|l}
\hline
                   &\#nodes    &\#edges &Ave. deg \\ \hline
TwPolitic         & 35k      & 274k    & 4.5\\ \hline
Chilean     & 127.4k      & 1150k    & 19\\ \hline
Covid         & 1124k      & 24062k    & 6\\ \hline
RedditNews         & 29k      & 1168k    & 22\\ \hline
WikiTalk         & 92.1k      & 360.8k    & 7\\ \hline
WikiElec         & 7.1k      & 107k    & 30\\ \hline
themarker         & 69.4k      & 1600k    & 47\\ \hline
\end{tabular}
\end{table}

\noindent\textbf{Pipelines}. We follow previous polarization detection pipelines~\cite{li2021unsupervised,fang2022polarized}: We assume no labeled data. The inputs are the graph structure $G(V,E)$ and the input feature matrix $X$. The goal is to cluster the nodes $V$ into two polarized classes. The evaluation metric is the percentage accuracy.

\noindent\textbf{Results}.

\begin{table*}[]
\caption {Clustering accuracy}
\centering
\label{table:exp}
\begin{tabular*}{\linewidth}{@{\extracolsep{\fill}} l|l|l|l|l|l|l|l}
\hline
                   &TwPolitic    &Chilean &Covid &RedditNews &WikiTalk &WikiElec & themarker \\ \hline
Grace          & \text{0.864}      & \text{0.793}    & \text{0.882} & \text{0.924} & \text{0.880} & \text{0.764} & \text{0.835} \\ \hline
CCA-SSG   & \text{0.882}      & \text{0.812}    & \text{0.895} & \text{0.916} & \text{0.880} & \text{0.751} & \text{0.841} \\ \hline
GraphMAE2    & \text{0.851}      & \text{0.820}    & \text{0.894} & \text{0.923} & \text{0.882} & \text{0.773} & \text{0.834} \\ \hline
P-GNN    & \text{0.855}      & \text{0.817}    & \text{0.864} & \text{0.909} & \text{0.894} & \text{0.769} & \text{0.851} \\ \hline
VGE    & \text{0.847}      & \text{0.798}    & \text{0.865} & \text{0.894} & \text{0.865} & \text{0.760} & \text{0.832} \\ \hline
FJ    & \text{0.809}      & \text{0.762}    & \text{0.805} & \text{0.884} & \text{0.865} & \text{0.722} & \text{0.800} \\ \hline
Hostile    & \text{0.798}      & \text{0.737}    & \text{0.724} & \text{0.911} & \text{0.767} & \text{0.695} & \text{0.792} \\ \hline
Patterns    & \text{0.812}      & \text{0.764}    & \text{0.817} & \text{0.901} & \text{0.807} & \text{0.807} & \text{0.804} \\ \hline
DocTra    & \textbf{0.906}      & \textbf{0.867}    & \textbf{0.923} & \textbf{0.932} & \textbf{0.902} & \textbf{0.833} & \textbf{0.864} \\ \hline
\end{tabular*}
\end{table*}

\noindent\textbf{Discussion}. Overall, the self-supervised methods (Grace, CCA-SSG, GraphMAE2, P-GNN) outperform classical polarized detection methods (FJ, VGE), demonstrating the effectiveness of contrastive objectives in graph pre-training. Although the self-supervised objectives are general-purpose, the contrastive principle enables robust embeddings able to well-distinguish graph nodes. The characteristic-specific methods (Hostile and Patterns) perform well in certain datasets that align with their designing principle but perform terribly in others.

\subsection{Semi-supervision}
This section evaluates performance with supervision. We consider two types of supervision: (1). Node labels: 1\%, 2\%, and 5\% nodes are labeled (2). Class initialization: input a noisy version of ground truth where 30\% and 60\% labels are corrupted. We only compare to self-supervised baselines as they are capable of utilizing supervision.

\begin{figure*}[!htb]
   \begin{minipage}{0.33\textwidth}
     \centering
     \includegraphics[width=1\linewidth]{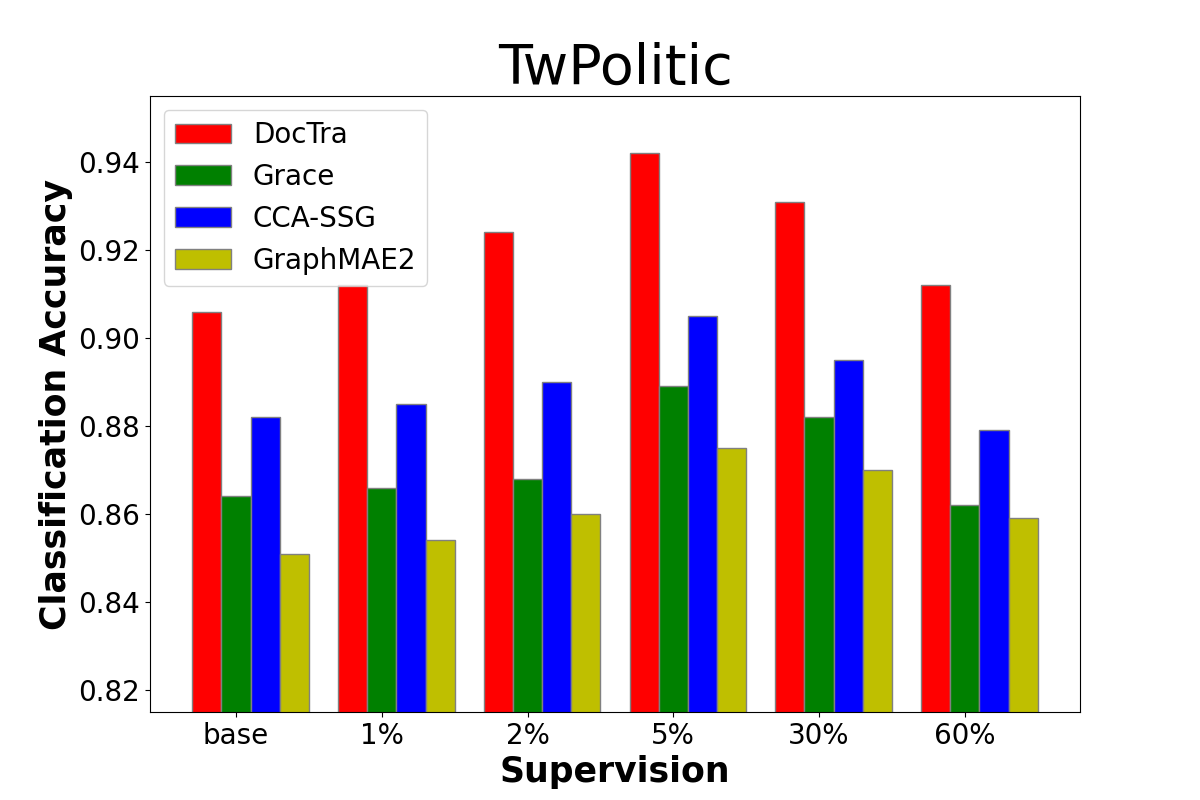}
     \label{Fig:Data1}
   \end{minipage}\hfill
   \begin{minipage}{0.33\textwidth}
     \centering
     \includegraphics[width=1\linewidth]{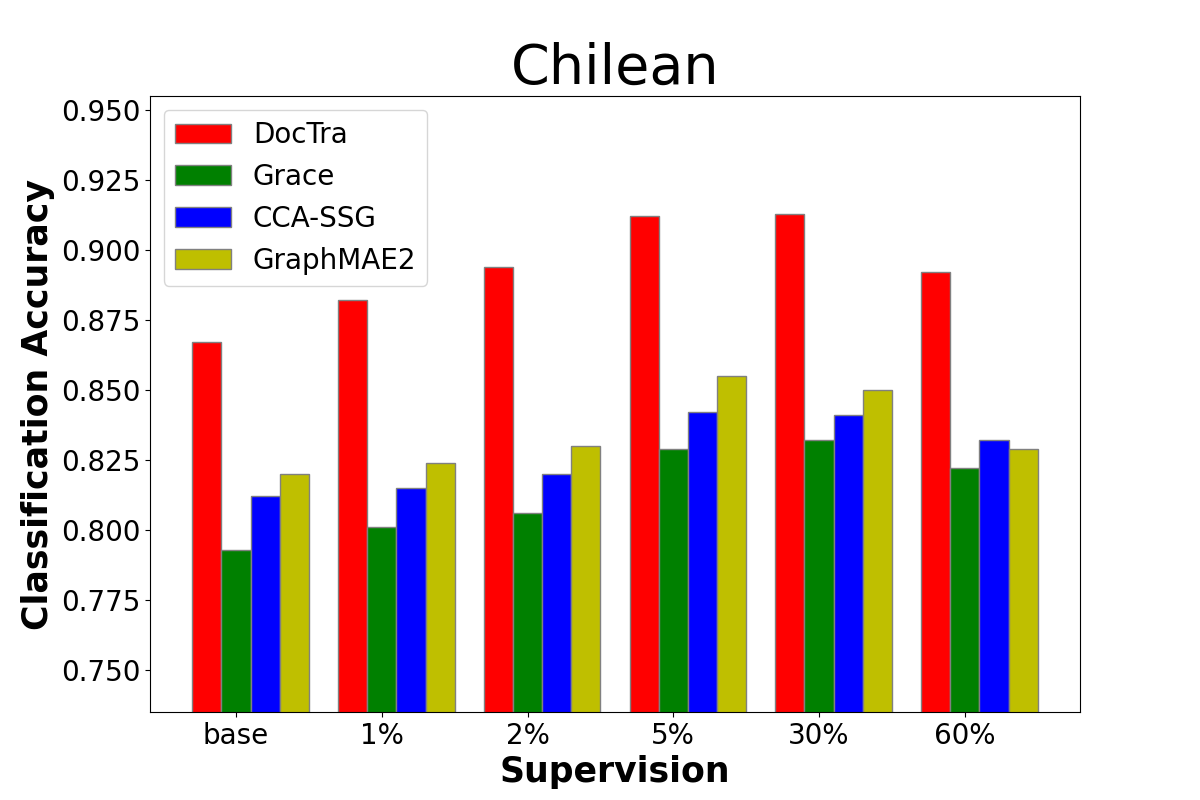}
    \label{Fig:Data2}
   \end{minipage}
   \begin{minipage}{0.33\textwidth}
     \centering
     \includegraphics[width=1\linewidth]{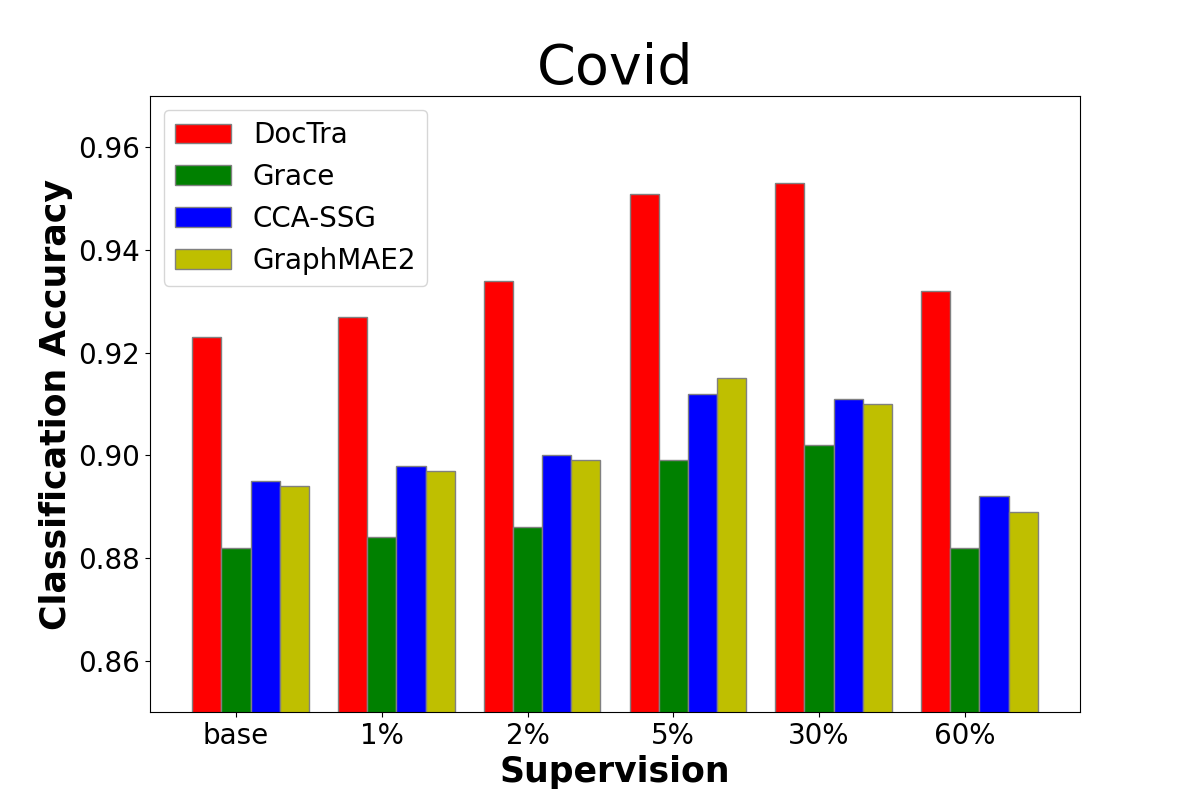}
     \label{Fig:Data1}
   \end{minipage}\hfill
   \begin{minipage}{0.33\textwidth}
     \centering
     \includegraphics[width=1\linewidth]{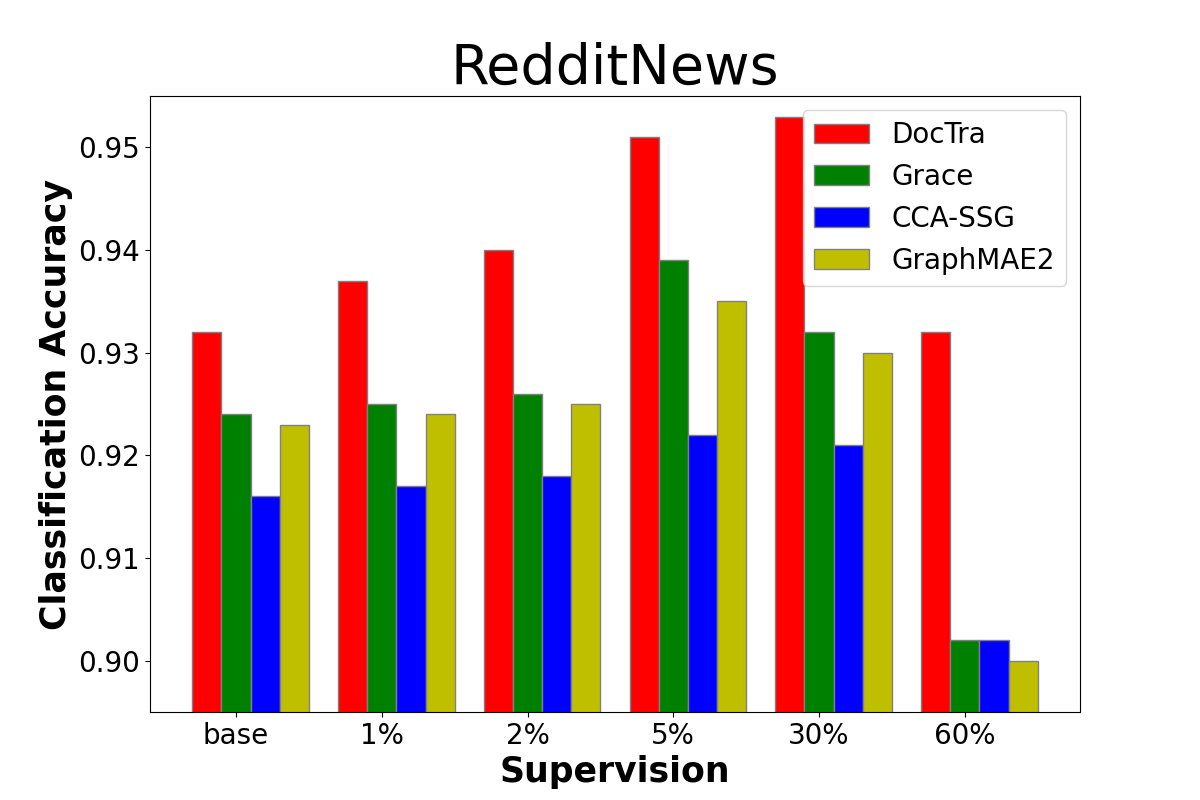}
    \label{Fig:Data2}
   \end{minipage}
   \begin{minipage}{0.33\textwidth}
     \centering
     \includegraphics[width=1\linewidth]{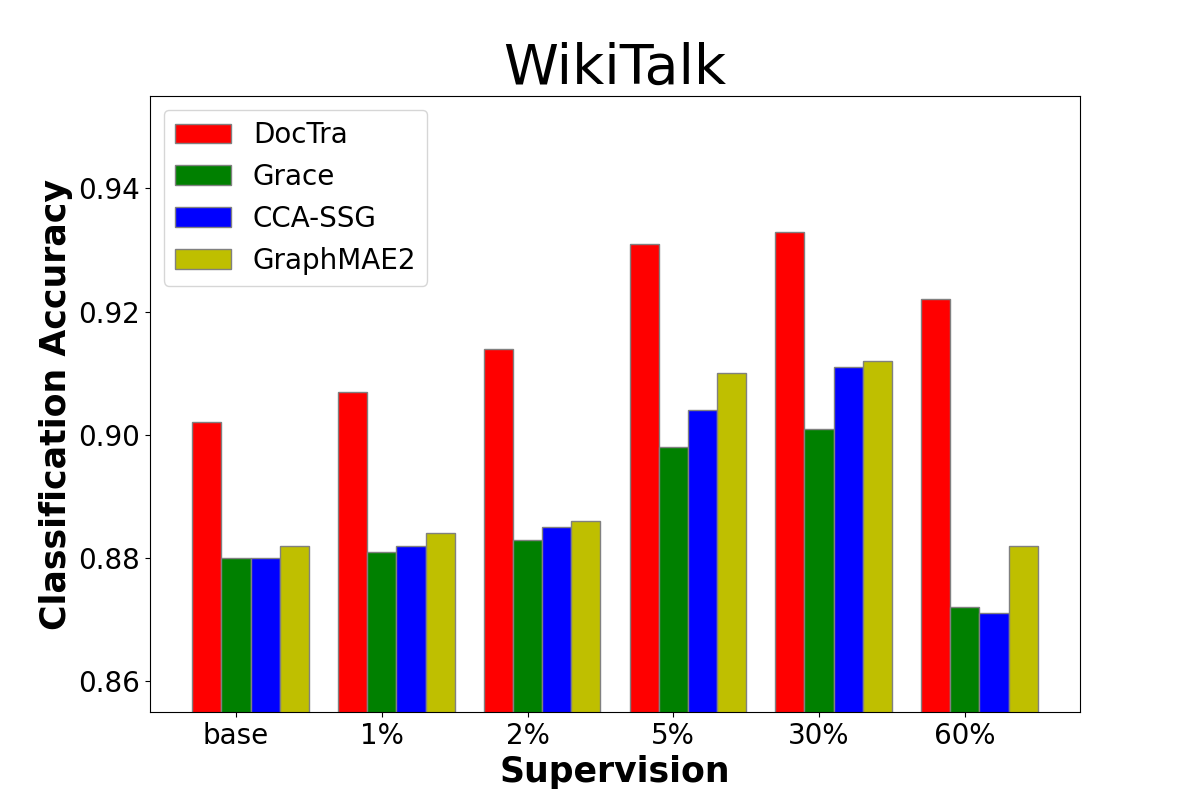}
     \label{Fig:Data1}
   \end{minipage}\hfill
   \begin{minipage}{0.33\textwidth}
     \centering
     \includegraphics[width=1\linewidth]{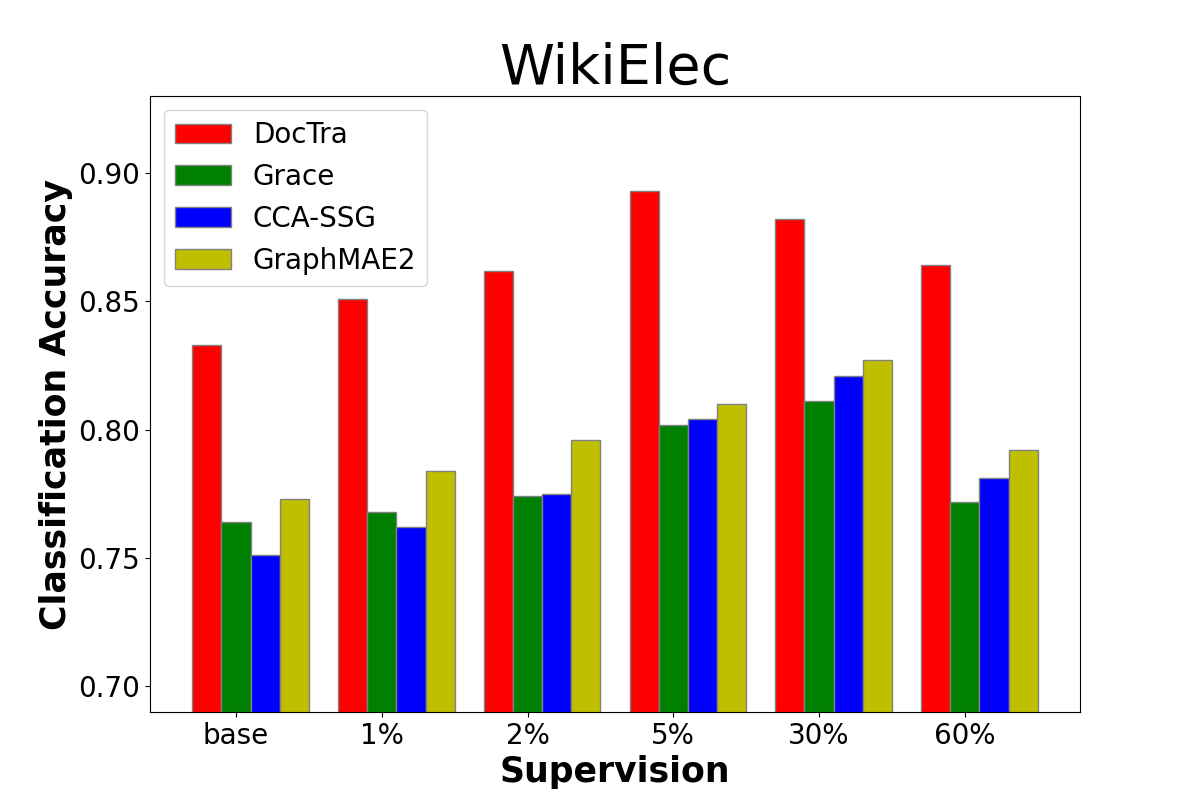}
    \label{Fig:Data2}
   \end{minipage}
   \caption{Polarization classification with semi-supervision}
   \label{fig:supervision}
\end{figure*}

The results are shown in fig.~\ref{fig:supervision}. In general, our Doctra benefits more from both supervised signals. The experiment suggests that $5\%$ labeled nodes is comparable to $30\%$ corrupted labels in polarization classification. Our method is also more robust to noise. In 60\% corrupted labels, our method overall gains performance while other baselines degrade.

\subsection{Unified Polarization Index}
This section evaluates the effectiveness of our proposed polarization metric in distinguishing polarized and unpolarized datasets. The level of polarization is often subjective and is hard to measure. Therefore, we pick several datasets that are universally recognized as not polarized in the literature to compare with the polarized datasets used in previous sections. The unpolarized datasets are \textit{Cora}, \textit{Citeseer}, \textit{PubMed}~\cite{yang2016revisiting}, \textit{Amazon-clothing}, and \textit{dblp}~\cite{kim2023task}. To compare our index with the \textit{polarization-disagreement index}, we normalize both into range $(0,1)$.

\begin{table}[]
\caption {Normalized Polarization Measurement}
\centering
\label{tab:index}
\begin{tabular}{l|l|l|l|l|l|l|l}
\hline
                   &TwPol    &Chilean &Covid &Reddit &WikiT &WikiE & themark \\ \hline
Our         & 0.82      & 0.80    & 0.81 & 0.79 & 0.85 & 0.77 & 0.82\\ \hline
p-d      & 0.78      & 0.66   & 0.61 & 0.79 & 0.66 & 0.63 & 0.72\\ \hline
                   &Cora    &Citeseer &PubM &Amaz &dblp \\ \hline
Our         & 0.22      & 0.17    & 0.31 & 0.29 & 0.25 \\ \hline
p-d      & 0.39      & 0.46   & 0.55 & 0.53 & 0.45\\ \hline
\end{tabular}
\end{table}

The results are shown in table.~\ref{tab:index}. Our unified polarization index is more effective as distinguishing the polarized datasets and unpolarized datasets. Notably, the traditional p-d index measures \textit{TwPolitic} and \textit{RedditNews} significantly more polarized than other datasets, which is not true. The underlying reason is that politics and news communities have higher background interaction levels. 

\subsection{Ablation Study}
This section performs an ablation study on our model by removing/replacing the essential building blocks, including contrastive objectives and $V^-_i$.

\begin{table*}[]
\small
\centering
\caption{Abalation Study}
\label{table:abalation}
\begin{tabular}{|c|c c c c c c c|}
\hline
 & TwPolitic  & Chilean & Covid & RedditNews & WikiTalk & WikiElec & themarker\\\hline
Base & $0.906$ & $0.867$ & $0.923$ & $0.932$ & $0.902$ & $0.833$ & $0.864$\\\hline
-$\mathcal{L}_i$ & $0.852$ & $0.821$ & $0.892$ & $0.901$ & $0.864$ & $0.793$ & $0.815$\\
-$\mathcal{L}_f$ & $0.882$ & $0.851$ & $0.906$ & $0.911$ & $0.874$ & $0.812$ & $0.834$\\\hline
$V^-_i$ & $0.854$ & $0.817$ & $0.862$ & $0.891$ & $0.876$ & $0.803$ & $0.826$\\
\hline
\end{tabular}
\end{table*}

The results are shown in Table.~\ref{table:abalation}. $\mathcal{L}_i$ has the biggest effect on performance since the interaction-level contrastive objective is the core objective distinguishing node interactions. $V^-_i$ also contributes to the performance as $V^-_i$ generates efficient contrastive pairs. $\mathcal{L}_f$ contributes the least but still demonstrates sufficient performance gains.

\section{Conclusion}
This paper presents dual contrastive objectives (DocTra) for polarization detection and clustering/classification. Our method is the first self-supervised learning scheme for polarization study and is flexible to various supervised signals. The dual contrastive objectives are interaction-level, which contrasts between positive and negative examples of interactions; and feature-level, which contrasts between polarized and invariant feature spaces. In addition, we propose a unified polarization index for polarization measurement of datasets, which enables automatic scaling to background engagements. Our experiments extensively evaluate our methods on 7 public datasets against 8 baselines.

\bibliographystyle{IEEEtran}
\bibliography{main}
\end{document}